# Microservice Maturity of Organizations

## Towards an assessment framework.

Jean-Philippe Gouigoux[1], Dalila Tamzalit[2], Joost Noppen[3]

[1]Group CTO SALVIA Développement, Bât. 270, 45, Av. Victor Hugo 93300 Aubervilliers, France
[2]Université de Nantes, CNRS, LS2N, F-44000, Nantes, France
[3] BT Applied Research, Adastral Park, Barrack Square, Martlesham, Ipswich IP5 3RE, UK
1jp.gouigoux@salviadeveloppement.com, 2Dalila.Tamzalit@univ-nantes.fr, 3 johannes.noppen@bt.com

**Abstract.** This early work aims to allow organizations to diagnose their capacity to properly adopt microservices through initial milestones of a Microservice Maturity Model (*MiMMo*). The objective is to prepare the way towards a general framework to help companies and industries to determine their microservices maturity. Organizations lean more and more on distributed web applications and Line of Business software. This is particularly relevant during the current Covid-19 crisis, where companies are even more challenged to offer their services online, targeting a very high level of responsiveness in the face of rapidly increasing and diverse demands. For this, microservices remain the most suitable delivery application architectural style. They allow agility not only on the technical application, as often considered, but on the enterprise architecture as a whole, influencing the actual financial business of the company. However, microservices adoption is highly risk-prone and complex. Before they establish an appropriate migration plan, first and foremost, companies must assess their degree of readiness to adopt microservices. For this, *MiMMo*, a Microservices Maturity Model framework assessment, is proposed to help companies assess their readiness for the microservice architectural style, based on their actual situation. *MiMMo* results from observations of and experience with about thirty organizations writing software. It conceptualizes and generalizes the progression paths they have followed to adopt microservices appropriately. Using the model, an organization can evaluate itself in two dimensions and five maturity levels and thus: (i) benchmark itself on its current use of microservices; (ii) project the next steps it needs to achieve a higher maturity level and (iii) analyze how it has evolved and maintain a global coherence between technical and business stakes.

## 1  Introduction and problem statement

With an increasingly connected world and the expectation of services being available online, organizations are increasingly faced with the challenge of delivering or use their software in a shape that can handle demand at scale and ready for the Cloud. This is much more relevant during the current crisis, where companies are even more challenged to offer their services online, targeting a very high level of responsiveness in the face of rapidly increasing and diverse demands.

In recent years, service-oriented architectures [7, 8, 9] have emerged as the most popular paradigm in this space, with in particular the concept of microservices [4], hyper-scalable small algorithms of a transactional nature, becoming one of the core building blocks to achieving these goals [10, 11]. The fine-grained nature of these microservices combined with their horizontal scaling properties allows companies to easily pivot their services while at the same time supporting large workloads inherent to modern online systems [4]. Microservices are a good targeted architecture for the modernization of software systems [12, 13]. However, for all the benefits microservice-based architecture offer, the journey for a company to migrate to this architectural style can be challenging and perilous. In fact, legacy processes and lack of knowledge are the main hurdles that companies face for adopting microservices [1, 2, 3]. In addition, several authors consider that microservices are not viable for every software system, as there are numerous trade-offs to consider [5, 6] and reasons to adopt microservices and how may vary considerably between different organizations.

In an ideal situation, an organization that delivers its software through this paradigm is fully aligned in both its technical and business parts. For example, from a technical point of view the organization has a comprehensive understanding of the size of its user base and the performance implications this has. The organization also understands which parts of the software architecture are most affected by this scale (e.g., a payment component in a webshop) and has isolated this to become one of the aforementioned microservices to support the scaling demand. In addition, the organization also has aligned its organizational structure to support this mode of development through hiring the right talent, empowering teams for rapid feature deployment and technology use, service-oriented earnings models, and having a fundamental understanding of the up- and downsides of the service-oriented paradigm across the organization. In this situation an organization is best placed to take full advantage of the benefits offered by a microservices architecture from both an economic and technical perspective while at the same time minimizing constraints and managing risks inherent to the approach. The organization in such situation can therefore be described as a *Mature Microservice Organization.*

However, the large majority of companies do not find themselves in the position described above. Many larger organizations are still in the early stages of the migration towards being primarily a software company and ensuring their software is of high quality and delivered at speed. Adopting microservices is much more than simply leveraging APIs as microservices, as, regrettably, many enterprises are understanding. Technical challenges in such migration paths typically include migrating legacy systems to scalable cloud architectures, defining and implementing software delivery pipelines, reimplementing software that in its original form does not support cloud-based deployment, etc. In addition, from an organizational perspective, the organization typically has to redefine its software development methods, upskill its staff and redefine how it makes money from the services it offers. For example, the company might have to shift from selling software wholesale to offering the functionality up as a service with a pay-as-you-go model for monetization. An organization who finds itself on this migration trajectory is less able to take maximum advantage of microservice-based software development and can therefore be classified as *Immature Microservice Organization.*

Naturally every organization will have completed a distinct set of challenges in this migration process, some focusing more on addressing legacy software and technical challenges first, while others emphasize change management of the business processes first. And while those steps do not necessarily make an organization mature, they can already enable the organization to gain initial benefits of microservice-based development. As such it can be argued that microservice maturity of an organization needs to be considered from several dimensions' points of view, like the usual business and technical dimensions.

The paper is organized as follow: this first section introduces the context and the problem statement. Section 2 presents *MiMMo*, the proposed Microservice Maturity Model. Section 3 illustrate the results of using the *MiMMo* through two organizations. Section 4 is dedicated to related works and Section 5 discuss the work and open some future research tracks to conclude.

## 2 MiMMo: towards a Microservice Maturity Assessment Framework

MiMMo proposes the initial milestones towards a general framework to help companies and industries determine their microservices maturity, one of the main challenges for organizations. It helps to determine at what stage of maturity they are before considering to build an appropriate strategy to adopt properly microservices with a substantial Return Over Investment.

### 2.1 General presentation

MiMMo proposes to consider two main dimensions of importance for companies: the *organizational dimension*, that supports the business strategy, and the *technical dimension*. In order to assess the degree of maturity, each dimension is declined in different levels of maturity. The maturity assessment obtained will represent a good starting point for organizations to evaluate the necessary effort to adopt microservices and if it is worth doing it. The strategy of adoption microservices can be thus built appropriately to the context of each organization.

The proposed MiMMo framework has been derived from authors' experience, mainly based on 6-years field observations of about thirty organizations, each having between 40 and 8000 users (**Table 1**[1]). By working with and advising these organizations, one of the industrial authors has observed the journeys of these organizations on their trajectory towards using microservices. By comparing journeys, identifying recurring patterns and successful actions taken, the authors have compiled their observations into a generalized framework that can be used by organizations to assess their microservices maturity and readiness to take this paradigm to the next level. This maturity assessment framework therefore is a heuristic advice mechanism that captures observed industry best practices. This paper presents the first steps.

---

[1] Names of companies are not given for aims of confidentiality but their type is specified.

### 2.2 Followed Methodology

The derivation of the Microservice Maturity Model (MiMMo) came in two main stages: *observations* and *elaboration*:

#### 2.2.1 Observations

This observations stage lasted from 2014 to 2019. It represents the period where one of the authors was working with and advising around thirty organizations. These studied organizations include public agencies, like ministries, regional and departmental councils mainly in France, as well as some large cities and some mid-sized private enterprises.

*Phase 1 – Implicit observations (2014-2018):* from 2014 to 2016, only public organizations have been addressed with projects of Information System alignment around a modular software suite. The evolution in time of maturity did not clearly appear as closely related with the technical aspect of Service Oriented Architecture, since the domains were so close (almost the same) and the organizations have the same public status. It was only logical that they behave and evolve the same way. In 2017, the approach of what started to get called a microservice architecture became a significant advantage for the organizations addressed and the board of the company decided to create a dedicated Business Unit to approach other categories of customers. New categories of companies started to be advised in 2018. The 20 companies concerned by this first phase are presented in **Table 1** – Left side.

**Phase 2 –** *Explicit observations (2018-2019)***:** the same step in evolution of the architecture appeared among the different customers. And although the method to help them changed radically from one to the other (in terms of length and mission content, but also in actors), the experience of the first phase helped to clearly identify a pattern of evolution. All companies that were accompanied, were they small (40 users) or big (8000 users), were they using one technology or another, were they operating on one business domain or some that were completely different, went through steps in their Information System evolution towards Service Orientation that were basically similar to each other, and also to the steps observed in the first phase. The 9 companies concerned by this second phase are presented in **Table 1** – Right side.

#### 2.2.2 Elaboration
Afterwards, in 2019, the same experience was shared between the authors, coming from different backgrounds, and since the evolution of maturity towards microservices architecture seemed to have common features whatever the context, the idea of an assessment model was devised among the authors. In 2020, based on their respective expertise, the authors analyzed notes, observations, and extrapolated and generalized observed best practices of successful transformational journeys. As a general methodology for derivation of MiMMo, a comparative analysis was performed of the experiences and best practices observed in the organizations by the authors. In particular shared successful behavior was identified, such as the application of enabling technologies and the restructuring of the organization with respect to new challenges in licensing. The identified patterns in turn were ordered chronologically based on observed change management to understand their logical progression. Finally, 5 stages were identified based on observable evolution steps inside organizations. The key thoughts are formalized within the MiMMo framework in terms of levels and dimension of maturity, detailed in Section 3.

Table 1. 29 Organizations observed during 6 years.

| 20 Observed organizations (2014-2018) | | | | 9 Observed organizations (2018-2019) | | | |
|---|---|---|---|---|---|---|---|
| Index | Type | Year | # Users | Domain | Index | Type | Year | # Users | Domain |

| Index | Type | Year | # Users | Domain |
|---|---|---|---|---|
| 1 | Regional Council | 2014 | 450 | Complete map |
| 2 | Departmental Council | 2014 | 200 | Persons + Finance |
| 3 | Regional Council | 2014 | 300 | IAM + Persons |
| 4 | Regional Council | 2014 | 500 | Persons |
| 5 | Regional Council | 2014 | 450 | Persons + Finance |
| 6 | Regional Council | 2014 | 300 | Persons |
| 7 | Software Editor | 2014 | 130 | Business Intelligence |
| 8 | Equipment Industry | 2014 | 8000 | Complete map |
| 9 | Regional Council | 2015 | 400 | Persons |
| 10 | Regional Council | 2015 | 650 | Persons + EDM + IAM |
| 11 | Regional Council | 2015 | 200 | Complete map |
| 12 | Chamber of Commerce | 2015 | 300 | Professionnal training |
| 13 | Departmental Council | 2016 | 80 | Complete map |
| 14 | Regional Council | 2016 | 800 | IAM |
| 15 | Regional Council | 2016 | 300 | Complete map |
| 16 | Equipment Industry | 2016 | 1200 | IAM |
| 17 | Software Editor | 2016 | 170 | Planning + Persons |
| 18 | City | 2017 | 120 | Persons |
| 19 | Chemical Industry | 2017 | 200 | Finance + BI |
| 20 | Regional Council | 2018 | 650 | IAM + Persons |
| 21 | Agriculture | 2018 | 1200 | Complete map |
| 22 | Pollutants management | 2018 | 900 | Complete map |
| 23 | Lawyers | 2018 | 200 | IAM + Persons |
| 24 | News industry | 2018 | 1500 | Geographical |
| 25 | Agriculture | 2019 | 4300 | Persons |
| 26 | Food transformation | 2019 | 1000 | Data ontology |
| 27 | Software Editor | 2019 | 800 | Persons |
| 28 | Software Editor | 2019 | 40 | Persons + IAM |
| 29 | Government agency | 2019 | 80 | Data ontology |

As this was a retrospective rather than an in-situ exercise, and therefore no systematic data collection and analysis could be performed, MiMMo was decided to be a heuristic assessment framework representative of observed industry experience, as a first step for a further research objective to form the foundation for a formalized maturity model.

## 3   Levels and Dimensions of the MiMMo Framework

The proposed Microservice Maturity Model (MiMMo) aims to capture the maturity of organizations for software delivery using microservices by identifying five distinct *levels of maturity* applied on *two dimensions*.

## 3.1 Maturity Levels

The authors identified 5 levels per dimension, totally based on observations from the field starting at level 1 (least mature) and up to level 5 (most mature). Each of these levels consists of a set of attributes and behaviors that can be observed in an organization that puts them at this level of maturity:

- **Level 1: Theory Understood but not Applied:** at this stage, the organization has received training and / or has gained a basic understanding of the microservices approach, but no effort has been taken in applying it. No projects have been defined to serve as a first application and no business plans have been put in place for the new methods of software delivery. The knowledge of microservices at this point is purely theoretical.
- **Level 2: Unskilled Application of Principles:** This second level starts when the first initiatives are taken to apply the new knowledge on microservices. Application projects start and the first microservices are created without a complete grasp of the implication of granularity nor the organizational implications. Generally, the first technical realizations are being delivered but the organization is not in a position to take advantage of the opportunities offered from a technical and business perspective.
- **Level 3: Microservices by the Book:** This is the level that should in theory be used as soon as the project starts and the team leaves level 1, but both the technical application and embedding in the organization and business model of microservices requires understanding them in the specific context of the organization. After acquiring this additional contextual knowledge at level 3 the organization is now capable of applying microservices by the book. Technically the software produced is sound and the organization has realigned its business model to take advantage of the new mode of software development and productization. At this level the organization can also be observed correcting some of the mistakes made at level 2, such as re-architecting software systems and redefining business roles.
- **Level 4: Expertise:** This level 4 corresponds to a level of maturity where the external principles recommendations have been digested by the team and are routinely used. Best practices are shared and teams not only follow them but understand what they stand for and what will be the risk of not following them. In some cases, adaptations are made to the best practices, but this is done with full consciousness of its impact to the software and the organization itself.
- **Level 5: Application Beyond Best Practice:** The highest level of maturity is considered reached when the principles of level 4 are constantly applied (not only a few best practices, but a major part of the state of the art on the domain) and the organization starts new best practices or advanced experimentation on its own. Generally, this level will be achieved on a few domains only where the organization has gained complete expertise and can now explore disruptive approaches to push the microservices benefits further.

**3.2 Maturity Dimensions**

The levels of maturity defined in the previous section can serve as a framework for organizations to assess how far along they are in the journey towards a full microservice-oriented organization. By determining their own behavior against the core elements described above, these organizations can determine their next steps to further their maturity. However, maturity with respect to microservices covers many different aspects of an organization. In addition to purely the technical challenges there are also organizational challenges, such as management buy-in and licensing models, and sustainability considerations, such as environmental and social sustainability concerns and implications for employees. To accommodate these specialized dimensions of maturity in this section, authors explore the most prominent maturity dimensions, i.e., technical and organizational maturity, and illustrate how the aforementioned maturity levels translate to these specific domains.

**3.3 Technical Maturity Dimension**

Most of the microservices adoption stories start from a grassroot experimentation from the developers and technical experts and, though many managers now know that the technical bits are not worth much when they are not accompanied with the right governance, there is an old craftsmanship reflex in IT that focuses attention for the information system on its technical implementation, since this is the easier part to observe changes on.

The global overview of the maturity-level applied on the technical dimension is given in **Fig. 1**. while **Table 2** gives examples and is an illustration of the maturity levels in the technical dimensions. It lists typical behaviors and practices observed depending on the level of maturity of the company on its technical path to microservices use. The correspondence between **Fig. 1**. and the correspondence matrix **Table 2** is made with corresponding letters (A), (B).… When there is no letter, that means that the observation is generic.

Even if some manifestation from one level can be observed at the same time as some from another level, it is rare that a company has a high level of maturity on one axis and a low one on another. In the end, the global technical maturity of the observed entity is an average of the possible behaviors listed below.

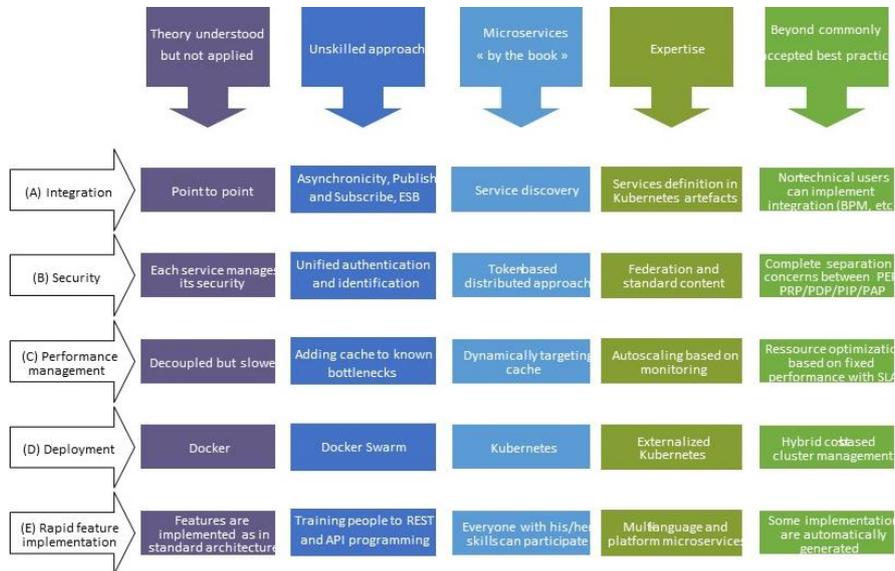

**Fig. 1.** Maturity-level correspondence matrix for the Technical Dimension.

**Table 2.** Maturity-level Correspondence Matrix Illustration for the Technical Dimension.

| Maturity levels | Technical Observations |
|---|---|
| Theory Understood but not Applied | • A few APIs have been coded but do not respect the REST best practices (there is a dedicated maturity model for RESTful APIs, developed by Leonard Richardson)<br>• (A) Integration between them or between them and the legacy systems is pure point to point, without any interfaces<br>• (B) The services deal with security in exactly the same way as the legacy system, without any contextual adaptation<br>• (C) First step of decoupling appears in the system, but this is mainly done at the cost of performance, since separation of responsibility is not compensated by adequate actions and the result is thus slower than the old monolith<br>• (D) These first tests are deployed with dedicated tools, or with Docker used in its simplest way, managing containers one by one<br>• (E) The features are developed using the same tools as for the legacy systems |
| Unskilled Approach | • Actual REST API are created and an external contract is created<br>• (A) A middleware is used to handle integration and calls between APIs, sometimes using publish and subscribe and sometimes direct calls<br>• (B) A unified authentication method is dedicated to microservices<br>• (C) Cache is added to restore performance of the whole, by choosing bottlenecks to correct<br>• (D) First orchestration approaches are attempted with Docker Swarm or other low-level techniques<br>• (E) Though the tools are not adapted to microservices, developers are trained to their specificities with respect to legacy code |
| Microservices by the Book | • APIs are created using the contract-first approach, and the definition of the API is handled by functional experts and not technical people anymore<br>• (A) A service discovery system is put in place<br>• (B) Authentication is standardized and uses a token-based approach to avoid performance impact of a central connected authentication service use<br>• (C) The performance of the services is monitored and dealt dynamically on each of them, while taking into account their functional dependencies upon each other's service level<br>• (D) Kubernetes or other high-level orchestration systems are put in place<br>• (E) Training has been achieved and developers start elaborate best practices on APIs and microservices. |
| Expertise | • (A) All services are contracted and exposed in a central directory based on the orchestration system<br>• (B) Federated security and external identity providers are routinely used for the microservices security<br>• (C) Performance is measured and adjusted continuously, depending on Service Level Agreement, current load and financial cost of the resources<br>• (D) The orchestrator is externalized or even hybrid, managing several cloud systems<br>• (E) Services are developed in the best platform for each usage, using the promise of the best tool for each service |
| Beyond Commonly Accepted Best Practices | • The whole company functions (not only the applications) is exposed in an API platform<br>• (A) Non-technical users are enabled to create value-added integration by plugging APIs together using dedicated middleware or low code platforms<br>• (B) Authentication, identification, authorization are completely isolated responsibilities<br>• (C) Resource use is balanced with performance for optimal usage depending on the business constraints solely<br>• (D) A multiple cloud system is used and the location of containers is fully hybrid, depending on cost and proximity to the source or the consumer of the service<br>• (E) Implementation of some low-level services are generated without any developer intervention |

### 3.3.1 Organizational Maturity Dimension.

As said previously, when considering microservice maturity it can be tempting to focus primarily on technical aspects, such as understanding of engineering principles and deployment. However, a second major component in the successful application of microservices is to have an organizational structure that is able to maximize the business potential of the benefits microservices have to offer. This not only includes providing development teams with access to relevant knowledge and technology, but also realignment of for example how software products are sold and which unique selling points these software products will have. The shift to microservices can lead to a realignment in target markets and even bring the company in competition with organizations it did not have to consider before. The global overview of the maturity-level applied on the organizational dimension is given in **Fig. 2.** is illustrated with **Table 3.**

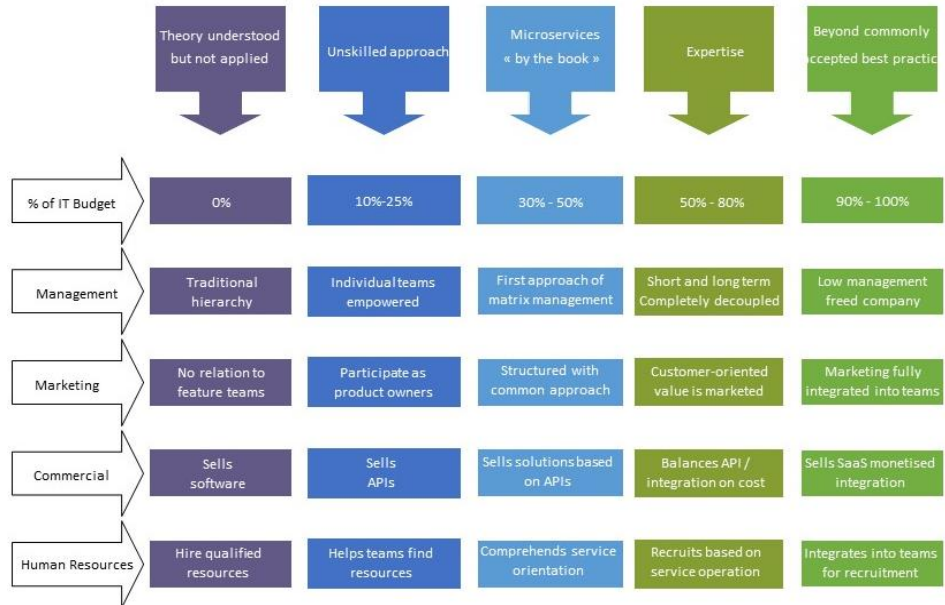

**Fig. 2.** Maturity-level correspondence matrix for the Organizational Dimension.

It is thus an illustration of the maturity levels for the organizational dimensions. It identifies behavior and attributes that can be observed in an organization at the corresponding level of maturity and will help the organization identify which steps need to be taken to drive the maturation process forward from an organizational perspective.

**Table 3.** Maturity-level Correspondence Matrix Illustration for the Organizational Dimension.

| Maturity levels | Organizational Observations |
|---|---|
| Theory Understood but not Applied | • The concept of microservices is understood uniformly across the organization<br>• Benefits and downsides can be discussed from a technical and business perspective without entrenched positions<br>• Implications for products offered and target market understood<br>• Implications for employees understood in terms of training and role definitions<br>• No decisions have been made and actions taken to initiate the creation and use of microservices in the organization |
| Unskilled Approach | • The organization has invested into creating their first microservices<br>• A small number of teams have gone through training and have been provided with tools and infrastructure<br>• Microservice versions of a small number of products are developed in parallel with existing software<br>• Market research is being done to determine how to monetize the new software products<br>• Information streams from technical teams to management to inform of product attributes, which needs to be translated to organizational change<br>• Initial disillusionment due to lack of expertise and early mistakes<br>• Recruitment of talent difficult due to a lack of understanding of what skillset is required |
| **Microservices by the Book** | • Having learned from the previous stage, technical and organizational blockers have been removed<br>• Relevant contracts for supporting developments have been agreed (e.g., cloud providers)<br>• Information stream between technical teams and management, with business requirements influencing development and technical knowledge influencing planning and feature strategy<br>• Talent recruitment is easier as a better appreciation of job expectations is put in place.<br>• Software licensing has shifted towards selling software as a service rather than a product<br>• Management has redefined its business model and future planning to center around microservices. |
| **Expertise** | • Substantial earnings and revenue of the organization is generated by microservice-based software<br>• Organization is comfortable expanding feature sets and moving into new markets, trusting their software to cope<br>• Development teams have been empowered to develop and restructure software systems based on technical considerations<br>• The organization explores and experiments with product and product features in rapid fashion, with minimal impact on workload and management<br>• New markets are now fully available to the organization and competition with technically capable competitors is possible on a consistent basis. |
| **Application Beyond Best Practice** | • The majority of software products offered by the company are now centered around microservice architectures<br>• Understanding of microservice benefits and limitations from a technical perspective and business implications permeate the entire organization<br>• Measures have been put in place to isolate the organization from vendor lock-in challenges<br>• Tooling support is explored and encouraged to ease development beyond core software teams and upskill relevant parts of the organization<br>• Active participation in microservices community is encouraged with employees being allowed to make in-house tools publicly available and attend and present at tech conferences<br>• Experimentation and sandboxing are supported and expected to be regular practice to ensure staying ahead of the curve<br>• Active knowledge sharing across the organization with a common sense of pride with respect to the level that has been reached<br>• Organizational image has changed towards being a high-tech company with customer trust in software offerings<br>• Able to compete with the best in the business and a business model that aims to achieve this<br>• Old software products (almost) completely phased out and maintenance no longer invested in |

As a result, the proposed model not only highlights the current maturity level but also a path to new attributes and behaviors that will make the organization as a whole better suited to microservice-based software delivery. It is good to note that these attributes and observations do not always apply uniformly and are not considered to be complete. However, they do highlight elements that have been observed in practice within companies with various levels of organizational maturity.

## 4   Illustration on two organizations

Out of 29 customers of one of the authors, two organizations illustrating very different trajectories in the maturity model have been chosen. The first one (for confidentiality reasons, let's call it *company A*) is a mid-sized established vertical software editor, with a strong technical culture. It is locally recognized as a pioneer in Service Oriented Architecture. The change to microservices was grassroot: the architects and technical leaders started the implementation as a full-blown replacement of the old monolithic architecture, while the financial and operational impacts were largely ignored. The maturity level on most of the technical axes is high and has reached target in most of them, while only some of the organizational axes have moved and some of them remain extremely low (**Table 4**) and radar diagram (**Fig. 3**). The use of the maturity model has helped in raising managers attention and internal as well as external training has been focused on functional- and commercial-oriented workforce. A remaining lack in integration and earning model still hurts financial return on the technical investment. Company A has been chosen because it represents, at its paramount level, the maturity path of a fair share of companies that have been observed by the authors. The corresponding maturity radar diagrams are established in **Fig. 3**.

The second one (for confidentiality reasons, let's call it *company B*) is a slightly larger but still middle-sized company that operates in retail services, also on a national scale. Though it does not belong to the software market, it could almost be considered as digital native, since most of its organization has been thought from the beginning around its information system. Company B has progressively come to a microservices approach, due to the same problems as company A, namely the increasingly problematic rate of evolution due to its monolithic information system. The main difference between the two companies, which are by other means quite comparable, is that company B has a financial approach to its software systems, and has prolonged the use of the legacy system until the risks indicators made it needed to think of a replacement. This replacement activity has been carefully thought of, based on a benchmark using the proposed maturity model in its technical dimension and evolution plan where costs, benefits and risks are modeled and adjusted along a three-year planning.

**Table 4.** Microservice Maturity Assessment of Company A.

| | | Objective | Achieved | Last eval |
|---|---|---|---|---|
| **Technical** | Integration | 5 | 2 | 2 |
| | Security | 4 | 3 | 3 |
| | Performance | 4 | 2 | 1 |
| | Deployment | 3 | 3 | 1 |
| | Rapid features | 3 | 2 | 2 |
| **Organizational** | Earning model | 3 | 1 | 0 |
| | Commercialization | 3 | 1 | 0 |
| | HR and support | 2 | 2 | 1 |
| | Knowledge management | 4 | 2 | 1 |
| | Business strategy | 4 | 0 | 0 |

**Fig. 3.** Maturity radar diagram of Company A.

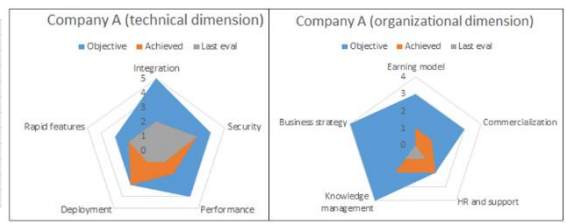

Though the objectives are high, the rate of evolution is more than satisfying. The drive on this plan, backed up by the managers, direction and even financial stakeholders is identified as the main reason for the high rate of transformation. Company B's maturity is stated through its assessment (**Table 5**).

**Table 5.** Microservice Maturity Assessment of Company B

| | | Company B | | |
|---|---|---|---|---|
| | | Objective | Achieved | Last eval |
| **Technical** | Integration | 3 | 2 | |
| | Security | 2 | 2 | |
| | Performance | 2 | 1 | |
| | Deployment | 4 | 1 | |
| | Rapid features | 2 | 2 | |
| **Organizational** | Earning model | 3 | 2 | |
| | Commercialization | 3 | 3 | |
| | HR and support | 4 | 3 | |
| | Knowledge management | 4 | 4 | |
| | Business strategy | 4 | 4 | |

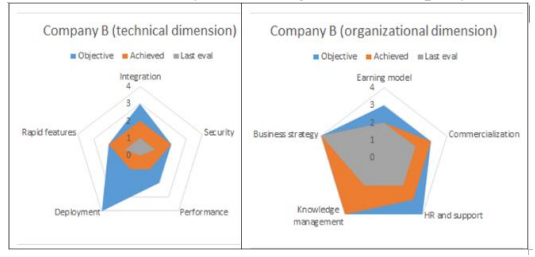

**Fig. 4.** Maturity radar diagram of Company B

Company B's maturity radar diagrams are represented above (**Fig. 4.**). It should be noted that these diagrams cannot be used to compare the rate of maturity change, but only the state of maturity itself, since the time between the last evaluation against the maturity model and the current evaluation that represents the achieved maturity is not the same for the two companies (more than two years for company A and less than a year for company B, which has started its microservices journey several years later but evolves much faster, certainly due to the fact that much more return of experience is now available).

## 5   Related Work

Maturity is a measurement of the ability of an organization for continuous improvement in a particular discipline [18]. Even if there is no consensus nor theoretical foundations on how to build them [19], there are multiple approaches for both researchers and practitioners to develop maturity models and a wide range of maturity assessment models have been developed as well by practitioners and academics over the past years. Almost each field (Analytics, Change Management, Continuous Delivery, Enterprise Architecture, Information Technology, Business Process Management…) has its proper maturity models. There are also universal maturity models like the most know of maturity models, Capability Maturity Model Integration (CMMI) [20]. The main idea of a maturity model is to briefly describe the typical behavior (activities) exhibited by an organization at a number of levels of maturity [21]. For each activity, it provides a description of the activity as it might be performed at each defined maturity level. Maturity models are designed to assess the maturity of a selected domain [22] and provides guidelines how to reach the next, higher maturity level [21]. All maturity models serve as informed approach for continuous improvement [20, 21] or as means of self or third-party assessment [25, 21]. MiMMo falls under the second category. It has been proposed to assess the current state and the desired future state of maturity of organizations regarding microservices. This paper proposes the initial milestones towards a general framework, with the objective to serve as a foundation for future research into a fully formalized maturity model, by involving as well academics and practitioners.

Even if there are reference architectures available in the service-oriented field, like the Open Group SOA Reference Architecture[2], there is no standard or reference architecture for micro-services. There are some beginning works, like the microservices capability model and a maturity model proposed in the book [26]. As for MiMMo, it is based on experience in industry. The capability model is divided in four areas: Core Capabilities (for components of a MS), Supporting Capabilities (not directly linked to MSs but to necessary for their development), Process & Governance Capabilities (tools and guidelines about MSs implementations) and Infrastructure Capabilities (for deployment and management of MSs). A maturity model presents 4 levels of maturity on 5 layers (application, database, infrastructure, monitoring and processes). It is interesting on the technical side but no organizational dimension is considered. In addition, no opening is considered to embody other dimensions.

From the academic view, regarding the microservice domain, several research works start to give some good pointers to the use of microservices and research trends. Among the most cited papers, Pahl & al. [15] proposed a mapping study and a characterization method but from the perspective of continuous development context, cloud and container technology but there is no organizational consideration. Jaramillo & al. [16] address leveraging microservices architecture via Docker technology. The book of Nadareishvili & al [17] addresses principles and practices of microservice architecture. These works are just an excerpt of numerous research contributions on the topic. However, the authors found only one contribution dedicated to the assessment of Microservices Maturity Models. Behara & al. [14] propose a Microservices Maturity Model. The paper outlines the problem of considering microservices by companies only from the technical dimension. They consider different assessment parameters (architecture, functional decomposition, codebase, data…) and, as for our MiMMo framework, different levels of maturity. They propose an assessment methodology but it is completely tied to the parameters they considered, making this methodology not applicable for any specific situation. According to the lack of research works on the topic and the important need of companies to assess their microservices maturity before considering their adoption and how, the authors considers that this field is in emergence. They thus proposed the first steps of MiMMo, that can continue to mature by leaning of all existing initiatives.

## 6 Discussions and Conclusion

### 6.1 Discussion

The MiMMo proposed in this paper aims to provide a framework and guidance for organizations who are keen to embark on and improve their use of microservices as the foundation for their software development. As MiMMo is a generalization of observed best-practice in industry across a large number of organizations, it will have general applicability. However, there are a number of challenges and discussion points that need to be considered.

---

[2] http://www.opengroup.org/soa/source-book/soa_refarch/index.htm

**Generalizability across Domains and Organizations:** while MiMMo is grounded in real-world observations of best practice in industry, it can be argued that further work is needed to establish its applicability across a wider range of domains. Depending on the starting point and knowledge inherent to the organization as well as the core business domain, refinement or adjustment of the model is required for aspects it currently does not consider. For example, it is highly likely that the organizations studied have gone through knowledge acquisition, training and experimentation even before making organizational changes. The current study does not consider such influences, making it a heuristic advice framework at this point rather than a full-fledged model. To ascertain completeness and to identify such refinement a more in-depth analysis and systematic evaluation of MiMMo is required. Another research track is to lean and to sustain the building of the model with the theory of design science.

**Longevity and Technology Progress:** a second consideration is the applicability of MiMMo in the long-term. Microservices are currently gaining in popularity, but their technology and management is rapidly changing and improving. In particular, the creation and management of microservices and serverless functions has been considerably streamlined over the last few years, which in turn will lower the bar for adoption of the technology. However, designing and architecting a software system and company infrastructure that is capable of taking advantage of this capability will remain as hard as before. It is likely that MiMMo will require continual updating with current best practices and deeper understanding of technical and business challenges to remain relevant. Further analysis of this is needed over a longer period of time to ensure the framework is up to date and has extension mechanisms that can cover these evolutions.

## 6.2   Conclusions

The main hurdles companies face for adopting microservices is lack of knowledge of how to adopt, in an appropriate way and according to their context, the microservice architectural style [1, 2, 3]. In this paper the authors proposed a Microservice Maturity Model (MiMMo) to help organizations to assess their degree of maturity in order to adopt the microservice architectural style by leaning before all on their situation, weakness and strengths. The proposed MiMMo represents the first milestones of an assessment framework upon which an organization can: (i) benchmark itself on its current use of microservices; (ii) project the next steps it needs to make in order to achieve a higher microservices maturity level and (iii) analyze how it has evolved and which area needs improvement to maintain global coherence between technical and business stakes.

The proposed MiMMo has been defined on the two most important dimensions of an organization: the technical and the organizational dimensions. Each of them has been declined in 5 levels of maturity in a correspondence matrix (**Fig. 1** and **Fig. 2**). Each of this matrix has been illustrated on encountered situations (**Table 2** and **Table 3**).

However, depending on the type of organization (for-profit, foundation, open-source, etc.) as well as the domain in which they are active, not only the technical and organizational dimensions can be refined, but likely additional dimensions can be identified and detailed as well the levels of maturity.

This paper presents the first steps, with the objective to serve as a foundation for future research into a fully formalized maturity model, by involving as well academics and organizations. For this, MiMMo is intended to be extensible by design, with the levels of maturity as a general categorization that is relevant and relatable across all dimensions. Adding a new dimension needs to establish the maturity-level correspondence matrix of the new dimension (for instance *Sustainability* dimension) by making the projection of the MiMMo level on the considered dimension, like in **Fig. 1** and **Fig. 2** for the Technical and Organizational dimensions. This creates the flexibility to define new and refined dimensions for specific domains and types of organizations in parallel to the dimensions already addressed in this article.

The overall maturity of an organization can then be interpreted as a combination of the score in each individual dimension, which can be represented for example as a score card or a spider chat. The authors did not propose a detailed methodology on how to assess the maturity of an organization with MiMMo. Proposing a methodology is not viable since there are so many numerous trade-offs to consider. Reasons to adopt microservices and how may vary considerably between different organizations [5, 6]. It completely depends on the organizations, their context and the adopted business strategy. The authors' position is to propose the MiMMo and explain its use through the case of two organizations and then let each organization find its proper barycenter of maturity. This is completely tied to the business strategy. Moreover, MiMMo is proposed based on authors experience. It needs to be improved in the future with interviews of industry experts to help practitioners develop their assessment capabilities, tied with the objective to facilitate academic contributions, ideally around a formal framework.